\documentclass[aps,twocolumn,showpacs,groupedaddress,amsmath,amssymb]{revtex4-1}

\usepackage{graphicx}
\usepackage{color}

\newcommand{\eqreff}[1]{Equation~(\ref{#1})}

\newcommand{\be}{\begin{equation}}
\newcommand{\ee}{\end{equation}}
\newcommand{\bes}{\begin{eqnarray}}
\newcommand{\ees}{\end{eqnarray}}
\newcommand{\bess}{\begin{eqnarray*}}
\newcommand{\eess}{\end{eqnarray*}}

\begin{document}

\title{
Fluctuation Theorem of Information Exchange within an Ensemble of Paths Conditioned on Correlated-Microstates
}

\author{Lee Jinwoo}
\email{e-mail: jinwoolee@kw.ac.kr}
\affiliation{Department of Mathematics, Kwangwoon 
University, 20 Kwangwoon-ro, Nowon-gu, Seoul 139-701, Korea}

\date{\today}

\begin{abstract}
Fluctuation theorems are a class of equalities \textcolor{black}{that express universal properties of the probability distribution of a fluctuating path functional such as heat, work or entropy production over an ensemble of trajectories during a non-equilibrium process with a well-defined initial distribution.} Jinwoo and Tanaka {(Jinwoo, L.;  Tanaka,  H.  \emph{Sci. Rep.} \textbf{2015}, \emph{5}, 7832)} 
have shown that work fluctuation theorems hold even within an ensemble of paths to each state, making it clear that entropy and free energy of each microstate encode heat and work, respectively, within the conditioned set. Here we show that information that is characterized by \textcolor{black}{the point-wise} mutual information for each {correlated} state between two subsystems in a heat bath encodes the entropy production of the subsystems and heat bath during a coupling process. To this end, we extend the fluctuation theorem of information exchange {(Sagawa, T.; Ueda,  M.  \emph{Phys. Rev. Lett.} \textbf{2012}, \emph{109}, 180602)} 
by showing that the fluctuation theorem holds even within an ensemble of paths that reach a {correlated} state during dynamic co-evolution of two subsystems.
\end{abstract}

\maketitle

\section{Introduction}

Thermal fluctuations play an important role in the functioning of molecular machines: fluctuations mediate the exchange of energy between molecules and the environment, enabling molecules to overcome free energy barriers and to stabilize in low free energy regions. They make positions and velocities random variables, and~thus make path functionals such as heat and work fluctuating quantities. In~the past two decades, a~class of relations called fluctuation theorems have shown that there are universal laws that regulate fluctuating quantities during a process that drives a system far from equilibrium. The~Jarzynski equality, for~example, links work to the change of equilibrium free energy~\cite{jar}, and~the Crooks fluctuation theorem relates the probability of work to the dissipation of work~\cite{crooks99} if we mention a few. There are many variations on these basic relations. Seifert has extended the second-law to the level of individual trajectories~\cite{seifert05}, and~Hatano and Sasa have considered transitions between steady states~\cite{sasa}. Experiments on single molecular levels have verified the fluctuation theorems, \textcolor{black}{providing critical insights on the behavior of bio-molecules~\cite{hummer,liph2001,liph2002,expSasa,expColin,ritort2012,PhysRevLett.121.120601,PhysRevLett.121.230601,PhysRevX.7.021051}.}

\textcolor{black}{Information is an essential subtopic of fluctuation theorems~\cite{PhysRevX.7.021003,yasar2014,yasar2018}. Beginning with pioneering studies on feedback controlled systems~\cite{PhysRevA_Lloyd,PhysRevE_Cao}, unifying formulations of information thermodynamics have been established~\cite{Gaspard_2013, PhysRevLett_Barato, PhysRevE_Barato, esposito2014prx, Rosinberg_2016}. 
Especially,} Sagawa and Ueda have introduced information to the realm of fluctuation theorems~\cite{sagawa}. They have established a fluctuation theorem of information exchange, unifying non-equilibrium processes of measurement and feedback control~\cite{sagawa2}. They have considered a situation where a system, say $X$, evolves in such a manner that depends on state $y$ of another system $Y$ the state of which is fixed during the evolution of the state of $X$. In~this setup, they have shown that establishing a correlation between the two subsystems accompanies an entropy production. Very~ recently, we have released the constraint that Sagawa and Ueda have assumed, and~proved that the same form of the fluctuation theorem of information exchange holds even when both subsystems $X$ and $Y$ co-evolve in time~\cite{jinwoo2019fluctuation}. 

In the context of fluctuation theorems, \textcolor{black}{external control $\lambda_t$ defines a process by varying the parameter in a predetermined manner during $0 \le t \le \tau$.} One repeats the process according to initial probability distribution $P_0$, and~then, a~system generates as a response an ensemble of microscopic trajectories $\{x_t\}$. Jinwoo and Tanaka~\cite{local,arxiv_v1} have shown that the Jarzynski equality and the Crooks fluctuation theorem hold even within an ensemble of trajectories conditioned \textcolor{black}{on} a fixed microstate at final time $\tau$, where the local form of non-equilibrium free energy replaces the role of equilibrium free energy \textcolor{black}{in the equations, making} it clear that free energy of microstate $x_\tau$ encodes the amount of supplied work for reaching $x_\tau$ during processes $\lambda_t$. Here local means that a term is related to microstate $x$ at time $\tau$ considered as an ensemble.

\begin{figure*}[t]
\centering
\includegraphics[width=15cm]{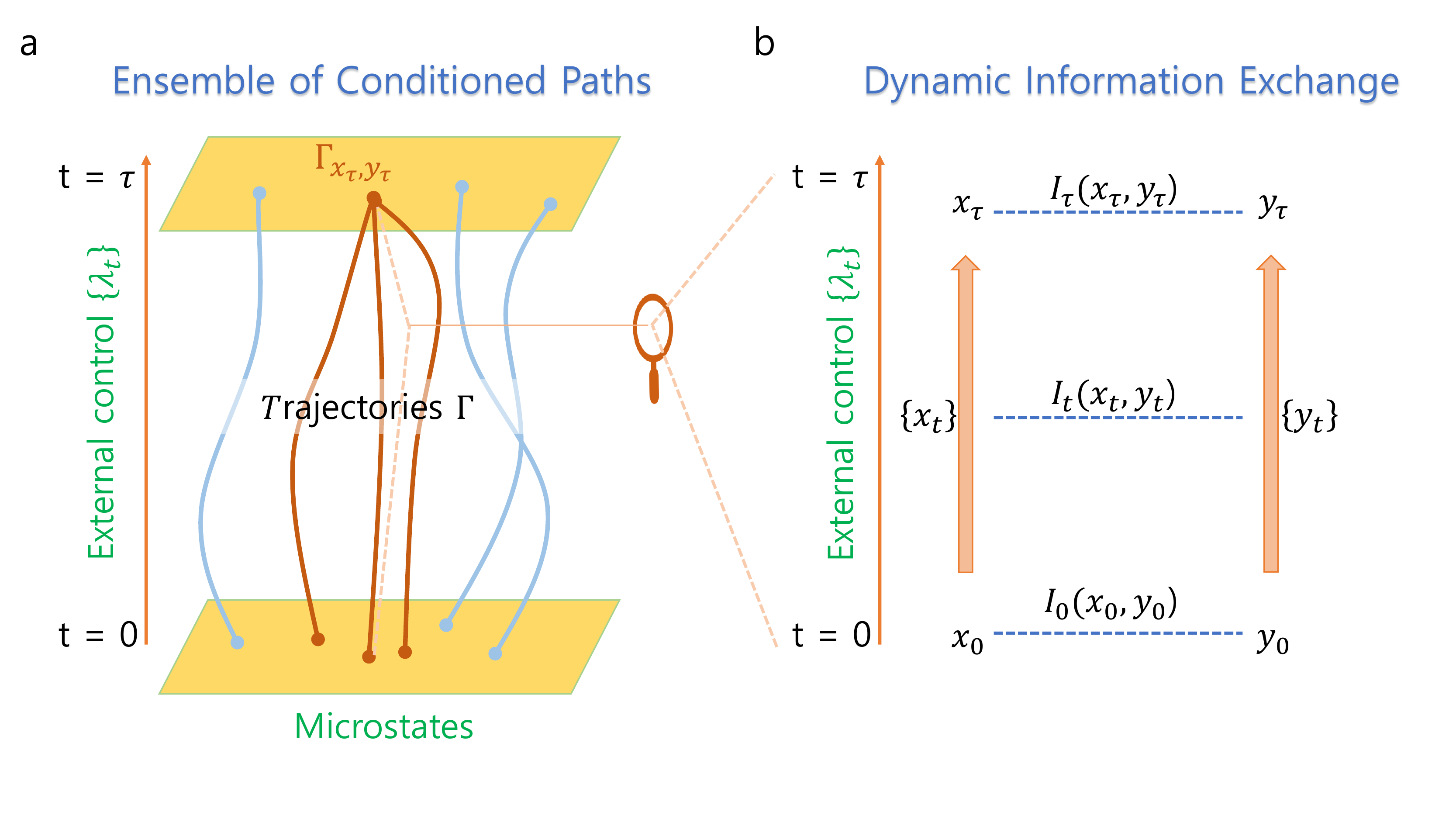}
\caption{Ensemble of conditioned paths and dynamic information exchange:
(\textbf{a}) \textcolor{black}{$\Gamma$ and $\Gamma_{x_\tau,y_\tau}$ denote respectively the~set of all trajectories during process $\lambda_t$ for $0\le t\le \tau$ and that of paths that reach $(x_\tau,y_\tau)$ at time $\tau$. Red curves schematically represent some members of $\Gamma_{x_\tau,y_\tau}$.} (\textbf{b}) We magnified a single trajectory in the left panel to represent a detailed view of dynamic coupling of \textcolor{black}{$(x_\tau, y_\tau)$ during process $\lambda_t$}. The \textcolor{black}{point-wise mutual information}~$I_t(x_t,y_t)$ may vary not necessarily monotonically.\label{fig1}}
\end{figure*}

In this paper, we apply this conceptual framework of considering a single microstate as an ensemble of trajectories to the fluctuation theorem of information exchange (see Figure~\ref{fig1}a). We show that mutual information of a {correlated}-microstates encodes the amount of entropy production within the ensemble of paths that reach the {correlated}-states. This local version of the fluctuation theorem of information exchange provides much more detailed information for each {correlated}-microstates compared to the results in~\cite{sagawa2, jinwoo2019fluctuation}. In~the existing approaches that consider the ensemble of all paths,  each point-wise mutual information does not provide specific details on a {correlated}-microstates, but~in this new approach of \textcolor{black}{focusing on} a subset of the ensemble, local mutual information provides detailed knowledge on particular {correlated}-states.

We organize the paper as follows: In Section~\ref{s2}, we briefly review some fluctuation theorems that we have mentioned. In~Section~\ref{s3}, we prove the main theorem and its corollary. In~Section~\ref{s4}, we provide illustrative examples, and~in Section~\ref{s5}, we discuss the implication of the~results. 

\section{Conditioned Nonequilibrium Work Relations and Sagawa--Ueda Fluctuation~Theorem}\label{s2}
We consider a system in contact with a heat bath of inverse temperate $\beta:=1/(k_BT)$ where $k_B$ is the Boltzmann constant, and~$T$ is the temperature of the heat bath. External parameter $\lambda_t$ drives the system away from equilibrium during $0 \le t \le \tau$. We assume that the initial probability distribution is equilibrium one at \textcolor{black}{control parameter} $\lambda_0$. Let $\Gamma$ be the set of all microscopic trajectories, and~$\Gamma_{x_\tau}$ be that of paths conditioned \textcolor{black}{on} $x_\tau$ at time $\tau$. Then, the~Jarzynski equality~\cite{jar} and end-point conditioned version~\cite{local,arxiv_v1} of it read as follows:
\bes\label{eq:jar}
F_{eq}(\lambda_\tau) &=& F_{eq}(\lambda_0)-\frac{1}{\beta}\ln \left< e^{-\beta W}\right>_\Gamma  \quad\mbox{and} \\
\label{eq:jinwoo}
{\cal F}(x_\tau,\tau) &=& F_{eq}(\lambda_0)-\frac{1}{\beta}\ln  \left< e^{-\beta W}\right>_{\Gamma_{x_\tau}},
\ees
respectively, where brackets $\left<\cdot\right>_\Gamma$ indicates the average over all trajectories in $\Gamma$ and $\left<\cdot\right>_{\Gamma_{x_\tau}}$ indicates the average over trajectories reaching $x_\tau$ at time $\tau$. 
Here $W$ indicates work done on the system through $\lambda_t$, $F_{eq}(\lambda_t)$ is equilibrium free energy at \textcolor{black}{control parameter} $\lambda_t$, and~${\cal F}(x_\tau,\tau)$ is local non-equilibrium free energy of $x_\tau$ at time $\tau$.
Work measurement over a specific ensemble of paths gives us equilibrium free energy as a function of $\lambda_\tau$ through \eqreff{eq:jar} and local non-equilibrium free energy as a micro-state function of $x_\tau$ at time $\tau$ through \eqreff{eq:jinwoo}. 
The following fluctuation theorem links Equations \eqref{eq:jar} and \eqref{eq:jinwoo}:
\be\label{eq:2}
\left<e^{-\beta {\cal F}(x_\tau,\tau)}\right>_{x_\tau} = e^{-\beta F_{eq}(\lambda_\tau)},
\ee
where brackets $\left<\cdot\right>_{x_\tau}$ indicates the average over all microstates $x_\tau$ at time $\tau$ \cite{local,arxiv_v1}.
Defining the reverse process by $\lambda'_t:=\lambda_{\tau-t}$ for $0 \le t \le \tau$, the~Crooks fluctuation theorem~\cite{crooks99} and end-point conditioned version~\cite{local,arxiv_v1} of it read as follows:
\bes\label{eq:crooks}
\frac{P_\Gamma(W)}{P'_\Gamma(-W)} &=&\exp\left(\frac{W-\Delta F_{eq}}{k_BT} \right) \quad \mbox{and}\\
\label{eq:jinwoo2}
\frac{P_{\Gamma_{x_\tau}}(W)}{P'_{\Gamma_{x_\tau}}(-W)} &=& \exp\left(\frac{W-\Delta {\cal F}}{k_BT} \right),
\ees
respectively, where $P_\Gamma(W)$ and  $P_{\Gamma_{x_\tau}}(W)$ are probability distributions of work $W$ normalized over all paths in $\Gamma$ and $\Gamma_{x_\tau}$, respectively. Here $P'$ indicates corresponding probabilities for the reverse process. For~\eqreff{eq:crooks}, the~initial probability distribution of the reverse process is an equilibrium one at \textcolor{black}{control parameter} $\lambda_\tau$. On~the other hand, for~\eqreff{eq:jinwoo2}, the~ initiail probability distribution for the reverse process should be \textcolor{black}{non-equilibrium probability distribution $p(x_\tau,\tau)$} of the forward process at \textcolor{black}{control parameter} $\lambda_\tau$. By~identifying such $W$ that $P_\Gamma(W) = P'_\Gamma(-W)$, one obtains $\Delta F_{eq}:=F_{eq}(\lambda_\tau)-F_{eq}(\lambda_0)$, the~difference in equilibrium free energy between $\lambda_0$ and $\lambda_\tau$, through \eqreff{eq:crooks} \cite{expColin}. Similar identification may provide $\Delta {\cal F}:={\cal F}(x_\tau,\tau)-F_{eq}(\lambda_0)$ through \eqreff{eq:jinwoo2}.

Now we turn to the Sagawa--Ueda fluctuation theorem of information exchange~\cite{sagawa2}. Specifically, we discuss the generalized version~\cite{jinwoo2019fluctuation} of it. To~this end, we consider two subsystems $X$ and $Y$ in the heat bath of inverse temperature $\beta$. 
During process $\lambda_t$, they interact and co-evolve with each other. Then, the~fluctuation theorem of information exchange reads as follows:
\be\label{eq:sagawa-jinwoo}
\left<e^{-\sigma+\Delta I}\right>_\Gamma = 1,
\ee
where brackets indicate the ensemble average over all paths of the combined subsystems, and~$\sigma$ is the sum of entropy production of system $X$, system $Y$, and~the heat bath, and~$\Delta I$ is the change in mutual information between $X$ and $Y$.
We note that in the original version of the Sagawa--Ueda fluctuation theorem, only system $X$ is in contact with the heat bath and $Y$ does not evolve during the process~\cite{sagawa2, jinwoo2019fluctuation}.
In this paper, we prove an end-point conditioned version of \eqreff{eq:sagawa-jinwoo}:
\be\label{eq:main_intro}
I_\tau(x_\tau,y_\tau) = - \ln \left<e^{-(\sigma+I_0)}\right>_{x_\tau,\, y_\tau},
\ee
where brackets indicate the ensemble average over all paths to $x_\tau$ and $y_\tau$ at time $\tau$, and~$I_t$ ($0 \le t \le \tau$) is local form of mutual information between microstates of $X$ and $Y$ at time $t$ (see Figure~\ref{fig1}b). If~there is no initial correlation, i.e.,~$I_0=0$, \eqreff{eq:main_intro} clearly indicates that local mutual information $I_\tau$ as a function of {correlated}-microstates $(x_\tau,y_\tau)$ encodes entropy production $\sigma$ within the end-point conditioned ensemble of paths. In~the same vein, we may interpret initial correlation $I_0$ as encoded entropy production for the preparation of the initial~condition.

\section{Results\label{s3}}

\subsection{Theoretical~Framework}
Let $X$ and $Y$ be finite classical stochastic systems in the heat bath of inverse temperate $\beta$. We allowed external parameter $\lambda_t$ drives one or both subsystems away from equilibrium during time $0 \le t \le \tau$ \cite{jarReview,revSeifert,review}. We assumed that classical stochastic dynamics describes the time evolution of $X$ and $Y$ during process $\lambda_t$ along trajectories $\{x_t\}$ and $\{y_t\}$, respectively, where $x_t$ ($y_t$) denotes a specific microstate of $X$ ($Y$) at time $t$ for $0\le t\le \tau$ on each trajectory.
Since trajectories fluctuate, we repeated process $\lambda_t$ with initial joint probability distribution $P_0(x,y)$ over all microstates $(x, y)$ of systems $X$ and $Y$. Then the subsystems may generate a joint probability distribution $P_t(x,y)$ for $0\le t\le \tau$. Let $P_t(x):=\int P_t(x,y)\,dy$ and $P_t(y):=\int P_t(x,y)\,dx$ be the corresponding marginal probability distributions.
We~assumed
\be\label{eq:neq}
P_0(x,y) \neq 0 \mbox{ for all } (x,y),
\ee
so that we have  
$P_t(x,y) \neq 0$, $P_t(x) \neq 0$, and~$P_t(y) \neq 0$ for all $x$ and $y$ during $0\le t \le\tau$.  
\textcolor{black}{Now we consider entropy production $\sigma$ of system $X$ along $\{x_t\}$, system $Y$ along $\{y_t\}$, and~heat bath $Q_b$ during process $\lambda_t$ for $0\le t\le \tau$ as follows}
\be\label{eq:sigma}
\sigma:=\Delta s  + \beta Q_b,
\ee 
where
\bes\label{eq:s}
\begin{aligned}
\Delta s &:= \Delta s_x + \Delta s_y,\\
\Delta s_x &:= -\ln P_\tau(x_\tau) + \ln P_0(x_0), \\
\Delta s_y &:= -\ln P_\tau(y_\tau) + \ln P_0(y_0).
\end{aligned}
\ees
\textcolor{black}{We remark that \eqreff{eq:s} is different from the change of stochastic entropy of combined super-system composed of $X$ and $Y$, which reads $\ln P_0(x_0,y_0) -\ln P_\tau(x_\tau,y_\tau)$ that reduces to \eqreff{eq:s} if processes $\{x_t\}$ and $\{y_t\}$ are independent. The~discrepancy leaves room for correlation \eqreff{eq:I} below~\cite{sagawa2}.} Here the stochastic entropy $s[P_t(\circ)]:=-\ln P_t(\circ)$ of microstate $\circ$ at time $t$ is uncertainty of $\circ$ at time $t$: the more uncertain that microstate $\circ$ occurs, the~greater the stochastic entropy of $\circ$ is.
We also note that in~\cite{sagawa2}, system $X$ was in contact with the heat reservoir, but~system $Y$ was not. Nor did system $Y$ evolve. Thus their entropy production reads $\sigma_{\rm su} := \Delta s_x + \beta Q_b$.

Now we assume, during~process $\lambda_t$, that system $X$ exchanged information with system $Y$. By~this, we mean that trajectory $\{x_t\}$ of system $X$ evolved depending on the trajectory $\{y_t\}$ of system $Y$ (see Figure~\ref{fig1}b). Then, the~local form of mutual information $I_t$ at time $t$ between $x_t$ and $y_t$ is the reduction of uncertainty of $x_t$ due to given $y_t$ \cite{sagawa2}: 
\bes\label{eq:I}
\begin{aligned}
I_t(x_t,y_t) &:= s[P_t(x_t)] - s[P_t(x_t | y_t)] \\
      &\,\, = \ln \frac{P_t(x_t, y_t)}{P_t(x_t)P_t(y_t)},
\end{aligned}
\ees
where $P_t(x_t|y_t)$ is the conditional probability distribution of $x_t$ given $y_t$. The~more information was being shared between $x_t$ and $y_t$ for their occurrence, the~larger the value of $I_t(x_t,y_t)$ was. 
We note that if $x_t$ and $y_t$ were independent at time $t$, $I_t(x_t,y_t)$ became zero. The~average of $I_t(x_t,y_t)$ with respect to $P_t(x_t,y_t)$ over all microstates is the mutual information between the two subsystems, which was greater than or equal to zero~\cite{cover}. 

\subsection{Proof of Fluctuation Theorem of Information Exchange Conditioned \textcolor{black}{on} a {Correlated}-Microstates}
Now we are ready to prove the fluctuation theorem of information exchange conditioned \textcolor{black}{on} a {correlated}-microstates. 
We define reverse process $\lambda'_t := \lambda_{\tau-t}$ for $0\le t\le \tau$, where the external parameter is time-reversed~\cite{ponmurugan2010,horowitz2010}. The~initial probability distribution $P'_0(x,y)$ for the reverse process should be the final probability distribution for the forward process $P_\tau(x,y)$ so that we~have 
\bes\label{eq:pp}
\begin{aligned}
P'_0(x)=\int P'_0(x,y)\, dy = \int P_\tau(x,y)\,dy = P_\tau(x), \\
P'_0(y)=\int P'_0(x,y)\, dx = \int P_\tau(x,y)\,dx = P_\tau(y). 
\end{aligned}
\ees
Then, by~  \eqreff{eq:neq}, we have $P'_t(x,y) \neq 0$, $P'_t(x) \neq 0$, and~$P'_t(y) \neq 0$ for all $x$ and $y$ during $0\le t \le\tau$. 
For each trajectories $\{x_t\}$ and $\{y_t\}$ for $0\le t\le \tau$, we define the time-reversed conjugate as~ follows:
\bes
\begin{aligned}
\{x'_t\}:=\{x^*_{\tau-t}\}, \\
\{y'_t\}:=\{y^*_{\tau-t}\},
\end{aligned}
\ees
where $*$ denotes momentum reversal.
Let $\Gamma$ be the set of all trajectories $\{x_t\}$ and $\{y_t\}$, and~$\Gamma_{x_\tau,y_\tau}$ be that of trajectories conditioned on {correlated}-microstates $(x_\tau, y_\tau)$ at time $\tau$. Due to time-reversal symmetry of the underlying microscopic dynamics, the~set $\Gamma'$ of all time-reversed trajectories was identical to $\Gamma$, and~the set $\Gamma'_{x'_0,y'_0}$ of time-reversed trajectories conditioned \textcolor{black}{on} $x'_0$ and $y'_0$ was identical to $\Gamma_{x_\tau,y_\tau}$. Thus we may use the same notation for both forward and backward pairs. We note that the path probabilities $P_\Gamma$ and $P_{\Gamma_{x_\tau,y_\tau}}$ were normalized over all paths in $\Gamma$ and $\Gamma_{x_\tau,y_\tau}$, respectively (see Figure~\ref{fig1}a).
With this notation, the~microscopic reversibility condition that enables us to connect the probability of forward and reverse paths to dissipated heat  reads as follows~\cite{kur, maes1999,crooks99,jar2000}:
\be\label{eq:Qb}
\frac{P_\Gamma(\{x_t\}, \{y_t\} | x_0,y_0)}{P'_\Gamma(\{x'_t\},\{y'_t\} | x'_0,y'_0)}=e^{\beta Q_b},
\ee
where $P_\Gamma(\{x_t\}, \{y_t\} | x_0,y_0)$ is the conditional joint probability distribution of paths $\{x_t\}$ and $\{y_t\}$ conditioned \textcolor{black}{on} initial microstates $x_0$ and $y_0$, and~$P'_\Gamma(\{x'_t\},\{y'_t\} | x'_0,y'_0)$ is that for the reverse process. 
Now we restrict our attention to those paths that are in $\Gamma_{x_\tau,y_\tau}$, and~divide both numerator and denominator of the left-hand side of \eqreff{eq:Qb} by $P_\tau(x_\tau,y_\tau)$. Since $P_\tau(x_\tau,y_\tau)$ is identical to $P'_0(x'_0,y'_0)$, \eqreff{eq:Qb} becomes as follows:
\bes\label{eq:Qb2}
\frac{P_{\Gamma_{x_\tau,y_\tau}}(\{x_t\},\{y_t\}|x_0,y_0)}{P'_{\Gamma_{x_\tau,y_\tau}}(\{x'_t\},\{y'_t\} | x'_0,y'_0 )} & = & e^{\beta Q_b} ,
\ees
since the probability of paths is now normalized over $\Gamma_{x_\tau,y_\tau}$.
Then we have the following:
\begin{widetext}
\bes\label{eq:l1}
\frac{P'_{\Gamma_{x_\tau,y_\tau}}(\{x'_t\},\{y'_t\} )}{P_{\Gamma_{x_\tau,y_\tau}}(\{x_t\},\{y_t\})} & = & 
\frac{P_{\Gamma_{x_\tau,y_\tau}}'(\{x'_t\},\{y'_t\} | x'_0,y'_0)}{P_{\Gamma_{x_\tau,y_\tau}}(\{x_t\}, \{y_t\} | x_0,y_0)}\cdot \frac{P'_0(x'_0,y'_0)}{P_0(x_0,y_0)} \\  \label{eq:l2}
& = & \frac{P'_{\Gamma_{x_\tau,y_\tau}}(\{x'_t\},\{y'_t\} | x'_0,y'_0)}{P_{\Gamma_{x_\tau,y_\tau}}(\{x_t\}, \{y_t\} | x_0,y_0)}\cdot\frac{P'_0(x'_0,y'_0)}{P'_0(x'_0)p'_0(y'_0)}\cdot \frac{P_0(x_0)P_0(y_0)}{P_0(x_0,y_0)} \\ \nonumber
& & \qquad\qquad\qquad \times \frac{P'_0(x'_0)}{P_0(x_0)}\cdot\frac{P'_0(y'_0)}{P_0(y_0)}\\ \label{eq:l3}
& = & \exp\{ - \beta Q_b + I_\tau(x_\tau,y_\tau) - I_0(x_0,y_0) - \Delta s_x - \Delta s_y\} \\ \label{eq:l4}
& = & \exp\{- \sigma + I_\tau(x_\tau,y_\tau) - I_0(x_0,y_0) \}.
\ees
\end{widetext}
To obtain \eqreff{eq:l2} from \eqreff{eq:l1}, we multiply \eqreff{eq:l1} by $\frac{P'_0(x'_0)P'_0(y'_0)}{P'_0(x'_0)P'_0(y'_0)}$ and $\frac{P_0(x_0)P_0(y_0)}{P_0(x_0)P_0(y_0)}$, which are $1$. We obtain \eqreff{eq:l3} by applying Equations \eqref{eq:s}--\eqref{eq:pp} and \eqref{eq:Qb2} to \eqreff{eq:l2}. Finally, we use \eqreff{eq:sigma} to obtain \eqreff{eq:l4} from \eqreff{eq:l3}.
Now we multiply both sides of \eqreff{eq:l4} by $e^{-I_\tau(x_\tau,y_\tau)}$ and $P_{\Gamma_{x_\tau,y_\tau}}(\{x_t\},\{y_t\})$, and~take integral over all paths in $\Gamma_{x_\tau,y_\tau}$ to obtain the fluctuation theorem of information exchange conditioned \textcolor{black}{on} a {correlated}-microstates:
\begin{widetext}
\bes\label{eq:main}
\begin{aligned}
\left<e^{-(\sigma+I_0)}\right>_{x_\tau,y_\tau} &:= \int_{\{x_t\},\{y_t\}\in\Gamma_{\{x_\tau\},\{y_\tau\}}} e^{-(\sigma+ I_0)} P_{\Gamma_{x_\tau,y_\tau}}(\{x_t\},\{y_t\}) \, d\{x_t\}d\{y_t\} \\
& = \int_{\{x_t\},\{y_t\}\in\Gamma_{\{x_\tau\},\{y_\tau\}}} e^{-I_\tau(x_\tau,y_\tau)}P'_{\Gamma_{x_\tau,y_\tau}}(\{x'_t\},\{y'_t\} ) \, d\{x'_t\}d\{y'_t\} \\
& = e^{-I_\tau(x_\tau,y_\tau)} \int_{\{x_t\},\{y_t\}\in\Gamma_{\{x_\tau\},\{y_\tau\}}} P'_{\Gamma_{x_\tau,y_\tau}}(\{x'_t\},\{y'_t\} ) \, d\{x'_t\}d\{y'_t\} \\
& = e^{-I_\tau(x_\tau,y_\tau)}.
\end{aligned}
\ees
\end{widetext}
Here we use the fact that $e^{-I_\tau(x_\tau,y_\tau)}$ is constant for all paths in $\Gamma_{x_\tau,y_\tau}$, probability distribution $P'_{\Gamma_{x_\tau,y_\tau}}$ is normalized over all paths in $\Gamma_{x_\tau,y_\tau}$, and~$d\{x_t\}=d\{x'_t\}$ and $d\{y_t\}=d\{y'_t\}$ due to the time-reversal symmetry~\cite{goldstein}. \eqreff{eq:main} clearly shows that just as local free energy encodes work~\cite{local}, and~local entropy encodes heat~\cite{arxiv_v1}, the~local form of mutual information between {correlated}-microstates $(x_\tau,y_\tau)$ encodes entropy production, within~the ensemble of paths that reach each microstate. 
The following corollary provides more information on entropy production in terms of energetic~costs.

\subsection{Corollary}

To discuss entropy production in terms of energetic costs, we define local free energy ${\cal F}_x$ of $x_t$ and ${\cal F}_y$ of $y_t$ at \textcolor{black}{control parameter} $\lambda_t$ as follows:
\bes\label{eq:psi}
\begin{aligned}
{\cal F}_x (x_t,  t) &:= E_x(x_t, t) - k_BTs[P_t(x_t)]  \\
{\cal F}_y (y_t,  t) &:= E_y(y_t, t) - k_BTs[P_t(y_t)],
\end{aligned}
\ees
where $T$ is the temperature of the heat bath, $k_B$ is the Boltzmann constant, $E_x$ and $E_y$ are internal energy of systems $X$ and $Y$, respectively, and~$s[P_t(\circ)]:=-\ln P_t(\circ)$ is stochastic entropy~\cite{crooks99,seifert05}. 
Work done on either one or both systems through process $\lambda_t$ is expressed by the first law of thermodynamics as follows:
\be\label{eq:1st}
W := \Delta E + Q_b,
\ee
where $\Delta E$ is the change in internal energy of the total system composed of $X$ and $Y$. 
If we assume that systems $X$ and $Y$ are weakly coupled, in~that interaction energy between $X$ and $Y$ is negligible compared to the internal energy of $X$ and $Y$, we may~have 
\be\label{eq:E}
\Delta E := \Delta E_x + \Delta E_y,
\ee
where $\Delta E_x := E_x(x_\tau,\tau) -E_x(x_0,0)$ and $\Delta E_y := E_y(y_\tau,\tau) -E_y(y_0,0)$ \cite{parrondo2015}.
We rewrite \eqreff{eq:l3} by adding and subtracting the change of internal energy $\Delta E_x$ of $X$ and $\Delta E_y$ of $Y$ as follows:
\begin{widetext}
\bes\label{eq:ll1}
\frac{P'_{\Gamma_{x_\tau,y_\tau}}(\{x'_t\},\{y'_t\} )}{P_{\Gamma_{x_\tau,y_\tau}}(\{x_t\},\{y_t\})}
& = & \exp\{ - \beta (Q_b + \Delta E_x + \Delta E_y) +  \beta \Delta E_x  - \Delta s_x + \beta \Delta E_y  - \Delta s_y\}\\ \nonumber
&&\qquad\qquad \times \exp\{ I_\tau(x_\tau,y_\tau) - I_0(x_0,y_0)  \}\\ \label{eq:ll2}
& = & \exp\{- \beta (W - \Delta {\cal F}_x - \Delta {\cal F}_y) + I_\tau(x_\tau,y_\tau) - I_0(x_0,y_0)   \},
\ees
\end{widetext}
where we have applied Equations \eqref{eq:psi}--\eqref{eq:E} consecutively to \eqreff{eq:ll1}  to obtain \eqreff{eq:ll2}. 
Here $\Delta {\cal F}_x:={\cal F}_x(x_\tau,\tau)-{\cal F}_x(x_0,0)$ and $\Delta{\cal F}_y := {\cal F}_y(y_\tau,\tau)-{\cal F}_y(y_0,0)$.
Now we multiply both sides of \eqreff{eq:ll2} by $e^{-I_\tau(x_\tau,y_\tau)}$ and $P_{\Gamma_{x_\tau,y_\tau}}(\{x_t\},\{y_t\})$, and~take integral over all paths in $\Gamma_{x_\tau,y_\tau}$ to obtain the following:
\begin{widetext}
\bes\label{eq:main2}
\begin{aligned}
\left<e^{- \beta (W - \Delta {\cal F}_x - \Delta {\cal F}_y) - I_0 } \right>_{x_\tau,y_\tau} 
&:= \int_{\{x_t\},\{y_t\}\in\Gamma_{\{x_\tau\},\{y_\tau\}}} e^{- \beta (W - \Delta {\cal F}_x - \Delta {\cal F}_y) - I_0   }  P_{\Gamma_{x_\tau,y_\tau}}(\{x_t\},\{y_t\}) \, d\{x_t\}d\{y_t\} \\
& = \int_{\{x_t\},\{y_t\}\in\Gamma_{\{x_\tau\},\{y_\tau\}}} e^{-I_\tau(x_\tau,y_\tau)}P'_{\Gamma_{x_\tau,y_\tau}}(\{x'_t\},\{y'_t\} ) \, d\{x'_t\}d\{y'_t\} \\
& = e^{-I_\tau(x_\tau,y_\tau)},
\end{aligned}
\ees
\end{widetext}
which generalizes known relations in the literature~\cite{kawai, takara, hasegawa, esposito2011, sagawa,parrondo2015}.
We note that \eqreff{eq:main2} holds under the weak-coupling assumption between systems $X$ and $Y$ during process $\lambda_t$, and~$\Delta {\cal F}_x + \Delta {\cal F}_y$ in \eqreff{eq:main2} is the difference in non-equilibrium free energy, which is different from the change in equilibrium free energy that appears in similar relations in the literature~\cite{kawai, takara, hasegawa, esposito2011, sagawa}.
If there is no initial correlation, i.e.,~$I_0=0$, \eqreff{eq:main2} indicates that local mutual information $I_\tau$ as a state function of {correlated}-microstates $(x_\tau,y_\tau)$ encodes entropy production, $\beta (W - \Delta {\cal F}_x - \Delta {\cal F}_y)$, within~the ensemble of paths in $\Gamma_{x_\tau,y_\tau}$. In~the same vein, we may interpret initial correlation $I_0$ as encoded entropy-production for the preparation of the initial~condition.

{In~\cite{sagawa2}, they showed that the entropy of $X$ can be decreased without any heat flow due to the negative mutual information change under the assumption that one of the two systems does not evolve in time. \eqreff{eq:main} implies that the negative mutual information change can decrease the entropy of $X$ and that of $Y$ simultaneously without any heat flow by the following:
\begin{equation}
\left<\Delta s_x+\Delta s_y \right>_{x_\tau,y_\tau} \ge \Delta I_\tau(x_\tau,y_\tau),
\end{equation} 
provided $\left<Q_b\right>_{x_\tau,y_\tau}=0$. Here $\Delta I_\tau(x_\tau,y_\tau):=I_\tau(x_\tau,y_\tau) -\left< I_0(x_0,y_0) \right>_{x_0,y_0}$.
In terms of energetics, \eqreff{eq:main2} implies that the negative mutual information change can increase the free energy of $X$ and that of $Y$ simultaneously without any external-supply of energy by the following:
\begin{equation}
 - \Delta I_\tau(x_\tau,y_\tau) \ge \beta\left<\Delta {\cal F}_x + \Delta {\cal F}_y\right>_{x_\tau,y_\tau} 
\end{equation}
provided $\left<W\right>_{x_\tau,y_\tau}=0$.
}

\begin{figure*}[t]
\centering
\includegraphics[width=15cm]{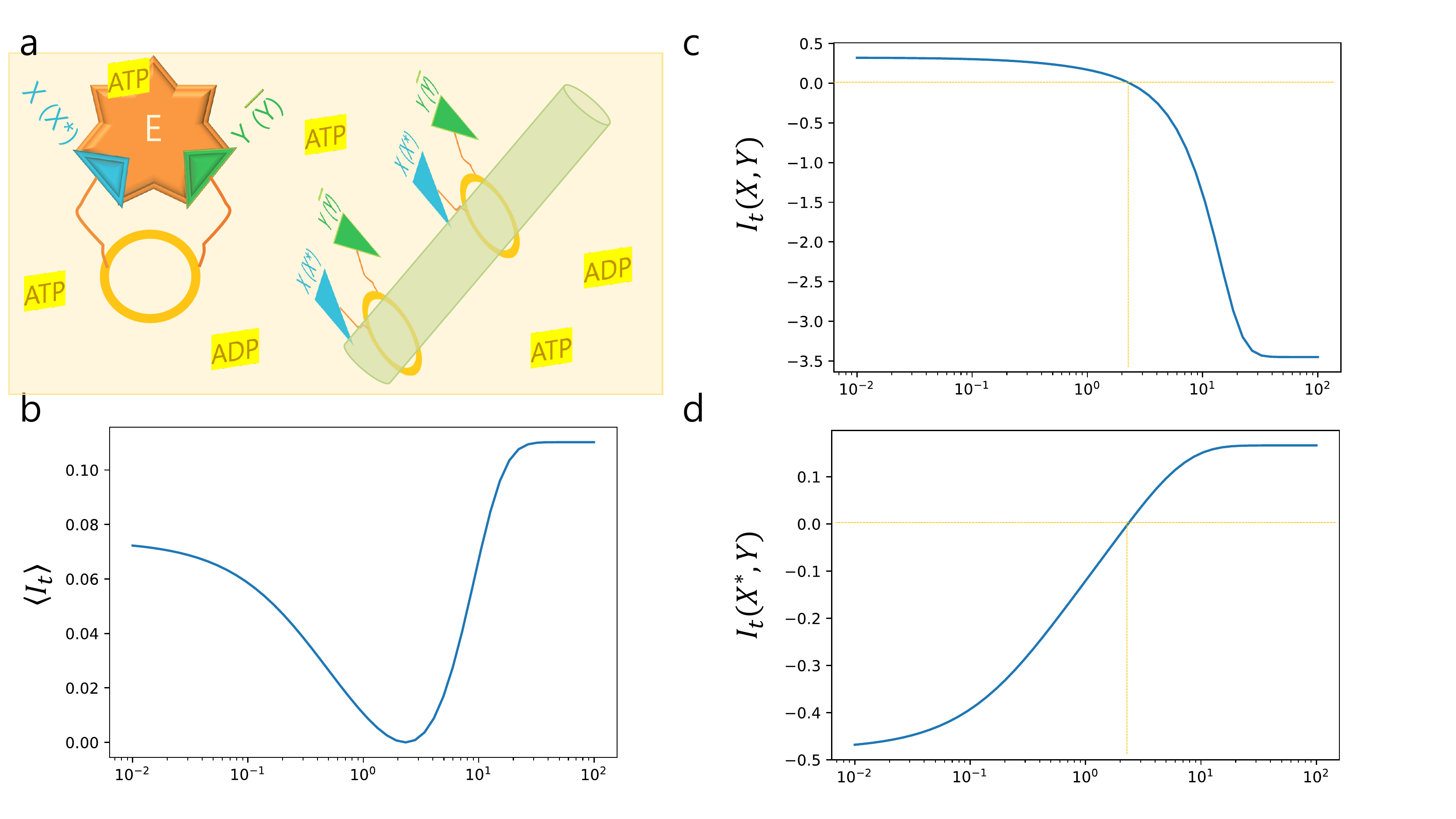}
\caption{\textcolor{black}{
Analysis of a ``tape-driven'' biochemical machine:
(\textbf{a}) a schematic illustration of enzyme $E$, pairs of $X (X^*)$ and $Y (\overline{Y})$ in the chemical bath including ATP and ADP. (\textbf{b}) The graph of $\left<I_t\right>_{\Gamma}$ as a function of time $t$, which shows the non-monotonicity of  $\left<I_t\right>_{\Gamma}$. (\textbf{c}) The graph of $I_t(X, Y)$ which decreases monotonically and composed of trajectories that harness mutual information to work \textcolor{black}{against} the chemical bath.  (\textbf{d}) The graph of $I_t(X^*, Y)$ that increases monotonically and composed of paths that create mutual information between $X^*$ and $Y$.}\label{fig2}}
\end{figure*}

\section{Examples\label{s4}}

\subsection{A Simple~One}

Let $X$ and $Y$ be two systems that weakly interact with each other, and~be in contact with the heat bath of inverse temperature $\beta$. We may think of $X$ and $Y$, for~example, as~bio-molecules that interact with each other or $X$ as a device which measures the state of other system and $Y$ be a measured system. We consider a dynamic coupling process as follows: Initially, $X$ and $Y$ are separately in equilibrium such that the initial correlation $I_0(x_0,y_0)$ is zero for all $x_0$ and $y_0$. At~time $t=0$, system $X$ {starts (weak) interaction with system $Y$ until time $t=\tau$}. During~the coupling process, external parameter $\lambda_t$ for $0\le t\le \tau$ may exchange work with either one or both systems (see Figure~\ref{fig1}b). Since each process fluctuates, we repeat the process many times to obtain probability distribution $P_t(x,y)$ for $0\le t\le \tau$. We allow both systems co-evolve interactively and thus $I_t(x_t,y_t)$ may vary not necessarily monotonically. Let us assume that the final probability distribution $P_\tau(x_\tau, y_\tau)$ is as shown in Table~\ref{tab1}.
\begin{table}[h]
\caption{The joint probability distribution of $x$ and $y$ at final time $\tau$: Here we assume that both systems $X$ and $Y$ have three states, $0$, $1$, and~$2$.\label{tab1}}
\centering
\begin{tabular}{cccc}
\toprule
\textbf{\boldmath$X\backslash Y$}	& \textbf{0}   & \textbf{1} & \textbf{2}  \\
\hline
\textbf{0}	 	& 1/6		& 1/9     &  1/18  \\
\textbf{1}		& 1/18  	& 1/6     &  1/9  \\
\textbf{2}           &  1/9       &  1/18   &  1/6\\
\hline
\end{tabular}
\end{table}
Then, a~few representative mutual information read as follows:
\bes\label{eq:I0}
\begin{aligned}
I_\tau(x_\tau=0,y_\tau=0) &= \ln\frac{1/6}{(1/3)\cdot(1/3)} = \ln (3/2), \\
I_\tau(x_\tau=0,y_\tau=1) &= \ln\frac{1/9}{(1/3)\cdot(1/3)} = 0, \\
I_\tau(x_\tau=0,y_\tau=2) &= \ln\frac{1/18}{(1/3)\cdot(1/3)} = \ln (1/2).
\end{aligned}
\ees
By Jensen's inequality~\cite{cover}, \eqreff{eq:main} implies 
\be\label{eq:coro}
\left<\sigma\right>_{x_\tau,y_\tau} \ge I_\tau(x_\tau,y_\tau).
\ee
Thus coupling $x_\tau=0, y_\tau=0$ accompanies on average entropy production of at least $\ln(3/2)$ which is greater than 0.
Coupling $x_\tau=0, y_\tau=1$ \textcolor{black}{may} not produce entropy on average.
Coupling $x_\tau=0, y_\tau=2$ on average may produce negative entropy by $\ln(1/2)=-\ln 2$. 
\textcolor{black}{Three individual inequalities provide more detailed information than that from $\left<\sigma\right>_\Gamma \ge \left< I_\tau(x_\tau,y_\tau)\right>_\Gamma \approx 0.0872$ currently available from~\cite{sagawa2,jinwoo2019fluctuation}.}

\subsection{A ``Tape-Driven'' Biochemical~Machine}

\textcolor{black}{In~\cite{Thomas2017}, McGrath~et~al. proposed a physically realizable device that exploits or creates mutual information, depending on system parameters. The~system is composed of an enzyme $E$ in a chemical bath, interacting with a tape that is decorated with a set of pairs of molecules (see Figure~\ref{fig2}a). A~pair is composed of substrate molecule $X$ (or phosphorylated $X^*$) and activator $Y$ of the enzyme (or $\overline{Y}$ which denotes the absence of $Y$). The~binding of molecule $Y$ to $E$ converts the enzyme into active mode $E^\dagger$, which catalyzes phosphate exchange between ATP and $X$:
\begin{equation}\label{eq:react}
X + {\rm ATP} + E^\dagger \rightleftharpoons E^\dagger\mbox{-}X\mbox{-}{\rm ADP}\mbox{-}P_i \rightleftharpoons 
E^\dagger + X^* + {\rm ADP}.
\end{equation} 
\indent The tape is prepared in a correlated manner through a single parameter $\Psi$:
\begin{eqnarray}
\begin{aligned}
p_0(\overline{Y}|X^*) &= p_0(Y|X)=\Psi, \\
p_0(Y | X^*) &= p_0(\overline{Y}|X)=1 - \Psi. 
\end{aligned}
\end{eqnarray}
\indent If $\Psi<0.5$, a~pair of $Y$ and $X^*$ is abundant so that the interaction of enzyme $E$ with molecule $Y$ activates the enzyme, causing the catalytic reaction of \eqreff{eq:react} from the right to the left, resulting in the production of ATP from ADP. If~the bath were prepared such that $[{\rm ATP}]>[{\rm ADP}]$, the~reaction corresponds to work on the chemical bath against the concentration gradient. Note that this interaction causes the conversion of $X^*$ to $X$, which reduces the initial correlation between $X^*$ and $Y$, resulting in the conversion of mutual information into work. If~$E$ interacts with a pair of $\overline{Y}$ and $X$ which is also abundant for $\Psi<0.5$, the~enzyme becomes inactive due to the absence of $Y$, preventing the reaction \eqreff{eq:react} from the left to the right, which plays as a ratchet that blocks the conversion of $X$ and ATP to $X^*$ and ADP, which might happen otherwise due to the the concentration gradient of the bath.
}

\textcolor{black}{On the other hand, if~$\Psi>0.5$, a~pair of $Y$ and $X$ is abundant which allows the enzyme to convert $X$ into $X^*$ using the pressure of the chemical bath, creating the correlation between $Y$ and $X^*$. If~$E$ interacts with a pair of $\overline{Y}$ and $X^*$ which is also abundant for $\Psi>0.5$, the~enzyme is again inactive, preventing the de-phosphorylation of $X^*$, keeping the created correlation. In~this regime, the~net effect is the conversion of work (due to the chemical gradient of the bath) to mutual information. The~concentration of ATP and ADP in the chemical bath is adjusted via $\alpha\in (-1,1)$ such that
\begin{equation}
[{\rm ATP}] = 1+\alpha \quad\mbox{ and }\quad [{\rm ADP}] = 1-\alpha
\end{equation}
relative to a reference concentration $C_0$. For~the analysis of various regimes of different parameters, we refer the reader to~\cite{Thomas2017}. 
}

\textcolor{black}{In this example, we concentrate on the case with $\alpha=0.99$ and $\Psi=0.69$, where Ref. \cite{Thomas2017} pays a special attention. They analyzed the dynamics of mutual information \textcolor{black}{$\left<I_t\right>_\Gamma$} during $10^{-2} \le t\le 10^{2}$.
Due to the high initial correlation, the~enzyme converts the mutual information between $X^*$ and $Y$ into work against the pressure of the chemical bath with $[{\rm ATP}]>[{\rm ADP}]$. As~the reactions proceed, correlation \textcolor{black}{$\left<I_t\right>_\Gamma$} drops until the minimum reaches, which is zero. Then, eventually the reaction is inverted, and~the bath begins with working to create mutual information between $X^*$ and $Y$ as shown in Figure~\ref{fig2}b. 
}

\textcolor{black}{We split the ensemble $\Gamma^t$ of paths into $\Gamma^t_{X,Y}$ composed of trajectories reaching $(X, Y)$ at each $t$ and $\Gamma^t_{X^*,Y}$ composed of those reaching $(X^*,Y)$ at time $t$. Then, we calculate $I_t(X, Y)$ and $I_t(X^*, Y)$ using the analytic form of probability distributions that they derived. Figure~\ref{fig2}c,d show $I_t(X, Y)$ and $I_t(X^*, Y)$, respectively, as~a function of time $t$. During~the whole process, mutual information $I_t(X, Y)$ monotonically decreases. For~$10^{-2}\le t \le 10^{1/3}$, it keeps positive, and~after that, it becomes negative which is possible for local mutual information. Trajectories in $\Gamma_{X,Y}$ harness mutual information between $X^*$ and $Y$, converting $X^*$ to $X$ and ADP to ATP against the chemical bath. Contrary to this, $I_t(X^*, Y)$ increases monotonically. It becomes positive after $t>10^{1/3}$, indicating that the members in $\Gamma^t_{X^*,Y}$ create mutual information between $X^*$ and $Y$ by converting $X$ to $X^*$ using the excess of $ATP$ in the chemical bath. The~effect accumulates, and~the negative values of $I_t(X^*, Y)$ turn to the positive after $t>10^{1/3}$.
}

\section{Conclusions\label{s5}}

We have proved the fluctuation theorem of information exchange conditioned \textcolor{black}{on}  {correlated}-microstates, \eqreff{eq:main}, and~its corollary, \eqreff{eq:main2}.
Those theorems make it clear that local mutual information encodes as a state function of {correlated}-states entropy production within an ensemble of paths that reach the {correlated}-states. \eqreff{eq:main} also reproduces lower bound of entropy production, \eqreff{eq:coro}, within~a subset of path-ensembles, which provides more detailed information than the fluctuation theorem involved in the ensemble of all paths. \eqreff{eq:main2} enables us to know the exact relationship between work, non-equilibrium free energy, and~mutual information. This end-point conditioned version of the theorem also provides more detailed information on the energetics for coupling than current approaches in the literature. This robust framework may be useful to analyze thermodynamics of dynamic molecular information processes~\cite{Becker2015, Thomas2017, Ouldridge2017} and to analyze dynamic allosteric transitions~\cite{tsai2014unified, cuendet2016allostery}.




\vspace{6pt} 




\noindent
{Acknowledgments.}
{L.J. was supported by the National Research Foundation of Korea Grant funded by the Korean Government (NRF-2010-0006733, NRF-2012R1A1A2042932, NRF-2016R1D1A1B02011106), and~in part by 
Kwangwoon University Research Grant in~2017.}



\bibliography{bib}

\begin{thebibliography}{48}%
\makeatletter
\providecommand \@ifxundefined [1]{%
 \@ifx{#1\undefined}
}%
\providecommand \@ifnum [1]{%
 \ifnum #1\expandafter \@firstoftwo
 \else \expandafter \@secondoftwo
 \fi
}%
\providecommand \@ifx [1]{%
 \ifx #1\expandafter \@firstoftwo
 \else \expandafter \@secondoftwo
 \fi
}%
\providecommand \natexlab [1]{#1}%
\providecommand \enquote  [1]{``#1''}%
\providecommand \bibnamefont  [1]{#1}%
\providecommand \bibfnamefont [1]{#1}%
\providecommand \citenamefont [1]{#1}%
\providecommand \href@noop [0]{\@secondoftwo}%
\providecommand \href [0]{\begingroup \@sanitize@url \@href}%
\providecommand \@href[1]{\@@startlink{#1}\@@href}%
\providecommand \@@href[1]{\endgroup#1\@@endlink}%
\providecommand \@sanitize@url [0]{\catcode `\\12\catcode `\$12\catcode
  `\&12\catcode `\#12\catcode `\^12\catcode `\_12\catcode `\%12\relax}%
\providecommand \@@startlink[1]{}%
\providecommand \@@endlink[0]{}%
\providecommand \url  [0]{\begingroup\@sanitize@url \@url }%
\providecommand \@url [1]{\endgroup\@href {#1}{\urlprefix }}%
\providecommand \urlprefix  [0]{URL }%
\providecommand \Eprint [0]{\href }%
\providecommand \doibase [0]{http://dx.doi.org/}%
\providecommand \selectlanguage [0]{\@gobble}%
\providecommand \bibinfo  [0]{\@secondoftwo}%
\providecommand \bibfield  [0]{\@secondoftwo}%
\providecommand \translation [1]{[#1]}%
\providecommand \BibitemOpen [0]{}%
\providecommand \bibitemStop [0]{}%
\providecommand \bibitemNoStop [0]{.\EOS\space}%
\providecommand \EOS [0]{\spacefactor3000\relax}%
\providecommand \BibitemShut  [1]{\csname bibitem#1\endcsname}%
\let\auto@bib@innerbib\@empty
\bibitem [{\citenamefont {Jarzynski}(1997)}]{jar}%
  \BibitemOpen
  \bibfield  {author} {\bibinfo {author} {\bibfnamefont {C.}~\bibnamefont
  {Jarzynski}},\ }\href@noop {} {\bibfield  {journal} {\bibinfo  {journal}
  {Phys. Rev. Lett.}\ }\textbf {\bibinfo {volume} {78}},\ \bibinfo {pages}
  {2690} (\bibinfo {year} {1997})}\BibitemShut {NoStop}%
\bibitem [{\citenamefont {Crooks}(1999)}]{crooks99}%
  \BibitemOpen
  \bibfield  {author} {\bibinfo {author} {\bibfnamefont {G.~E.}\ \bibnamefont
  {Crooks}},\ }\href@noop {} {\bibfield  {journal} {\bibinfo  {journal} {Phys.
  Rev. E}\ }\textbf {\bibinfo {volume} {60}},\ \bibinfo {pages} {2721}
  (\bibinfo {year} {1999})}\BibitemShut {NoStop}%
\bibitem [{\citenamefont {Seifert}(2005)}]{seifert05}%
  \BibitemOpen
  \bibfield  {author} {\bibinfo {author} {\bibfnamefont {U.}~\bibnamefont
  {Seifert}},\ }\href@noop {} {\bibfield  {journal} {\bibinfo  {journal} {Phys.
  Rev. Lett.}\ }\textbf {\bibinfo {volume} {95}},\ \bibinfo {pages} {040602}
  (\bibinfo {year} {2005})}\BibitemShut {NoStop}%
\bibitem [{\citenamefont {Hatano}\ and\ \citenamefont {Sasa}(2001)}]{sasa}%
  \BibitemOpen
  \bibfield  {author} {\bibinfo {author} {\bibfnamefont {T.}~\bibnamefont
  {Hatano}}\ and\ \bibinfo {author} {\bibfnamefont {S.-i.}\ \bibnamefont
  {Sasa}},\ }\href@noop {} {\bibfield  {journal} {\bibinfo  {journal} {Phys.
  Rev. Lett.}\ }\textbf {\bibinfo {volume} {86}},\ \bibinfo {pages} {3463}
  (\bibinfo {year} {2001})}\BibitemShut {NoStop}%
\bibitem [{\citenamefont {Hummer}\ and\ \citenamefont {Szabo}(2001)}]{hummer}%
  \BibitemOpen
  \bibfield  {author} {\bibinfo {author} {\bibfnamefont {G.}~\bibnamefont
  {Hummer}}\ and\ \bibinfo {author} {\bibfnamefont {A.}~\bibnamefont {Szabo}},\
  }\href@noop {} {\bibfield  {journal} {\bibinfo  {journal} {Proc. Nat. Acad.
  Sci. USA}\ }\textbf {\bibinfo {volume} {98}},\ \bibinfo {pages} {3658}
  (\bibinfo {year} {2001})}\BibitemShut {NoStop}%
\bibitem [{\citenamefont {Liphardt}\ \emph {et~al.}(2001)\citenamefont
  {Liphardt}, \citenamefont {Onoa}, \citenamefont {Smith}, \citenamefont
  {Tinoco},\ and\ \citenamefont {Bustamante}}]{liph2001}%
  \BibitemOpen
  \bibfield  {author} {\bibinfo {author} {\bibfnamefont {J.}~\bibnamefont
  {Liphardt}}, \bibinfo {author} {\bibfnamefont {B.}~\bibnamefont {Onoa}},
  \bibinfo {author} {\bibfnamefont {S.~B.}\ \bibnamefont {Smith}}, \bibinfo
  {author} {\bibfnamefont {I.}~\bibnamefont {Tinoco}}, \ and\ \bibinfo {author}
  {\bibfnamefont {C.}~\bibnamefont {Bustamante}},\ }\href@noop {} {\bibfield
  {journal} {\bibinfo  {journal} {Science}\ }\textbf {\bibinfo {volume}
  {292}},\ \bibinfo {pages} {733} (\bibinfo {year} {2001})}\BibitemShut
  {NoStop}%
\bibitem [{\citenamefont {Liphardt}\ \emph {et~al.}(2002)\citenamefont
  {Liphardt}, \citenamefont {Dumont}, \citenamefont {Smith}, \citenamefont
  {Tinoco~Jr},\ and\ \citenamefont {Bustamante}}]{liph2002}%
  \BibitemOpen
  \bibfield  {author} {\bibinfo {author} {\bibfnamefont {J.}~\bibnamefont
  {Liphardt}}, \bibinfo {author} {\bibfnamefont {S.}~\bibnamefont {Dumont}},
  \bibinfo {author} {\bibfnamefont {S.}~\bibnamefont {Smith}}, \bibinfo
  {author} {\bibfnamefont {I.}~\bibnamefont {Tinoco~Jr}}, \ and\ \bibinfo
  {author} {\bibfnamefont {C.}~\bibnamefont {Bustamante}},\ }\href@noop {}
  {\bibfield  {journal} {\bibinfo  {journal} {Science}\ }\textbf {\bibinfo
  {volume} {296}},\ \bibinfo {pages} {1832} (\bibinfo {year}
  {2002})}\BibitemShut {NoStop}%
\bibitem [{\citenamefont {Trepagnier}\ \emph {et~al.}(2004)\citenamefont
  {Trepagnier}, \citenamefont {Jarzynski}, \citenamefont {Ritort},
  \citenamefont {Crooks}, \citenamefont {Bustamante},\ and\ \citenamefont
  {Liphardt}}]{expSasa}%
  \BibitemOpen
  \bibfield  {author} {\bibinfo {author} {\bibfnamefont {E.~H.}\ \bibnamefont
  {Trepagnier}}, \bibinfo {author} {\bibfnamefont {C.}~\bibnamefont
  {Jarzynski}}, \bibinfo {author} {\bibfnamefont {F.}~\bibnamefont {Ritort}},
  \bibinfo {author} {\bibfnamefont {G.~E.}\ \bibnamefont {Crooks}}, \bibinfo
  {author} {\bibfnamefont {C.~J.}\ \bibnamefont {Bustamante}}, \ and\ \bibinfo
  {author} {\bibfnamefont {J.}~\bibnamefont {Liphardt}},\ }\href@noop {}
  {\bibfield  {journal} {\bibinfo  {journal} {Proc. Nat. Acad. Sci. USA}\
  }\textbf {\bibinfo {volume} {101}},\ \bibinfo {pages} {15038} (\bibinfo
  {year} {2004})}\BibitemShut {NoStop}%
\bibitem [{\citenamefont {Collin}\ \emph {et~al.}(2005)\citenamefont {Collin},
  \citenamefont {Ritort}, \citenamefont {Jarzynski}, \citenamefont {Smith},
  \citenamefont {Tinoco},\ and\ \citenamefont {Bustamante}}]{expColin}%
  \BibitemOpen
  \bibfield  {author} {\bibinfo {author} {\bibfnamefont {D.}~\bibnamefont
  {Collin}}, \bibinfo {author} {\bibfnamefont {F.}~\bibnamefont {Ritort}},
  \bibinfo {author} {\bibfnamefont {C.}~\bibnamefont {Jarzynski}}, \bibinfo
  {author} {\bibfnamefont {S.~B.}\ \bibnamefont {Smith}}, \bibinfo {author}
  {\bibfnamefont {I.}~\bibnamefont {Tinoco}}, \ and\ \bibinfo {author}
  {\bibfnamefont {C.}~\bibnamefont {Bustamante}},\ }\href@noop {} {\bibfield
  {journal} {\bibinfo  {journal} {Nature}\ }\textbf {\bibinfo {volume} {437}},\
  \bibinfo {pages} {231} (\bibinfo {year} {2005})}\BibitemShut {NoStop}%
\bibitem [{\citenamefont {Alemany}\ \emph {et~al.}(2012)\citenamefont
  {Alemany}, \citenamefont {Mossa}, \citenamefont {Junier},\ and\ \citenamefont
  {Ritort}}]{ritort2012}%
  \BibitemOpen
  \bibfield  {author} {\bibinfo {author} {\bibfnamefont {A.}~\bibnamefont
  {Alemany}}, \bibinfo {author} {\bibfnamefont {A.}~\bibnamefont {Mossa}},
  \bibinfo {author} {\bibfnamefont {I.}~\bibnamefont {Junier}}, \ and\ \bibinfo
  {author} {\bibfnamefont {F.}~\bibnamefont {Ritort}},\ }\href@noop {}
  {\bibfield  {journal} {\bibinfo  {journal} {Nature Phys.}\ } (\bibinfo {year}
  {2012})}\BibitemShut {NoStop}%
\bibitem [{\citenamefont {Holubec}\ and\ \citenamefont
  {Ryabov}(2018)}]{PhysRevLett.121.120601}%
  \BibitemOpen
  \bibfield  {author} {\bibinfo {author} {\bibfnamefont {V.}~\bibnamefont
  {Holubec}}\ and\ \bibinfo {author} {\bibfnamefont {A.}~\bibnamefont
  {Ryabov}},\ }\href {\doibase 10.1103/PhysRevLett.121.120601} {\bibfield
  {journal} {\bibinfo  {journal} {Phys. Rev. Lett.}\ }\textbf {\bibinfo
  {volume} {121}},\ \bibinfo {pages} {120601} (\bibinfo {year}
  {2018})}\BibitemShut {NoStop}%
\bibitem [{\citenamefont {\ifmmode~\check{S}\else \v{S}\fi{}iler}\ \emph
  {et~al.}(2018)\citenamefont {\ifmmode~\check{S}\else \v{S}\fi{}iler},
  \citenamefont {Ornigotti}, \citenamefont {Brzobohat\'y}, \citenamefont
  {J\'akl}, \citenamefont {Ryabov}, \citenamefont {Holubec}, \citenamefont
  {Zem\'anek},\ and\ \citenamefont {Filip}}]{PhysRevLett.121.230601}%
  \BibitemOpen
  \bibfield  {author} {\bibinfo {author} {\bibfnamefont {M.}~\bibnamefont
  {\ifmmode~\check{S}\else \v{S}\fi{}iler}}, \bibinfo {author} {\bibfnamefont
  {L.}~\bibnamefont {Ornigotti}}, \bibinfo {author} {\bibfnamefont
  {O.}~\bibnamefont {Brzobohat\'y}}, \bibinfo {author} {\bibfnamefont
  {P.}~\bibnamefont {J\'akl}}, \bibinfo {author} {\bibfnamefont
  {A.}~\bibnamefont {Ryabov}}, \bibinfo {author} {\bibfnamefont
  {V.}~\bibnamefont {Holubec}}, \bibinfo {author} {\bibfnamefont
  {P.}~\bibnamefont {Zem\'anek}}, \ and\ \bibinfo {author} {\bibfnamefont
  {R.}~\bibnamefont {Filip}},\ }\href {\doibase 10.1103/PhysRevLett.121.230601}
  {\bibfield  {journal} {\bibinfo  {journal} {Phys. Rev. Lett.}\ }\textbf
  {\bibinfo {volume} {121}},\ \bibinfo {pages} {230601} (\bibinfo {year}
  {2018})}\BibitemShut {NoStop}%
\bibitem [{\citenamefont {Ciliberto}(2017)}]{PhysRevX.7.021051}%
  \BibitemOpen
  \bibfield  {author} {\bibinfo {author} {\bibfnamefont {S.}~\bibnamefont
  {Ciliberto}},\ }\href {\doibase 10.1103/PhysRevX.7.021051} {\bibfield
  {journal} {\bibinfo  {journal} {Phys. Rev. X}\ }\textbf {\bibinfo {volume}
  {7}},\ \bibinfo {pages} {021051} (\bibinfo {year} {2017})}\BibitemShut
  {NoStop}%
\bibitem [{\citenamefont {Strasberg}\ \emph {et~al.}(2017)\citenamefont
  {Strasberg}, \citenamefont {Schaller}, \citenamefont {Brandes},\ and\
  \citenamefont {Esposito}}]{PhysRevX.7.021003}%
  \BibitemOpen
  \bibfield  {author} {\bibinfo {author} {\bibfnamefont {P.}~\bibnamefont
  {Strasberg}}, \bibinfo {author} {\bibfnamefont {G.}~\bibnamefont {Schaller}},
  \bibinfo {author} {\bibfnamefont {T.}~\bibnamefont {Brandes}}, \ and\
  \bibinfo {author} {\bibfnamefont {M.}~\bibnamefont {Esposito}},\ }\href
  {\doibase 10.1103/PhysRevX.7.021003} {\bibfield  {journal} {\bibinfo
  {journal} {Phys. Rev. X}\ }\textbf {\bibinfo {volume} {7}},\ \bibinfo {pages}
  {021003} (\bibinfo {year} {2017})}\BibitemShut {NoStop}%
\bibitem [{\citenamefont {Demirel}(2014)}]{yasar2014}%
  \BibitemOpen
  \bibfield  {author} {\bibinfo {author} {\bibfnamefont {Y.}~\bibnamefont
  {Demirel}},\ }\href@noop {} {\bibfield  {journal} {\bibinfo  {journal}
  {Entropy}\ }\textbf {\bibinfo {volume} {16}},\ \bibinfo {pages} {1931}
  (\bibinfo {year} {2014})}\BibitemShut {NoStop}%
\bibitem [{\citenamefont {Demirel}(2019)}]{yasar2018}%
  \BibitemOpen
  \bibfield  {author} {\bibinfo {author} {\bibfnamefont {Y.}~\bibnamefont
  {Demirel}},\ }\href@noop {} {\emph {\bibinfo {title} {Nonequilibrium
  Thermodynamics: Transport and Rate Processes in Physical, Chemical and
  Biological Systems}}}\ (\bibinfo  {publisher} {4th Ed., Elsevier,
  Amsterdam},\ \bibinfo {year} {2019})\BibitemShut {NoStop}%
\bibitem [{\citenamefont {Lloyd}(1989)}]{PhysRevA_Lloyd}%
  \BibitemOpen
  \bibfield  {author} {\bibinfo {author} {\bibfnamefont {S.}~\bibnamefont
  {Lloyd}},\ }\href {\doibase 10.1103/PhysRevA.39.5378} {\bibfield  {journal}
  {\bibinfo  {journal} {Phys. Rev. A}\ }\textbf {\bibinfo {volume} {39}},\
  \bibinfo {pages} {5378} (\bibinfo {year} {1989})}\BibitemShut {NoStop}%
\bibitem [{\citenamefont {Cao}\ and\ \citenamefont
  {Feito}(2009)}]{PhysRevE_Cao}%
  \BibitemOpen
  \bibfield  {author} {\bibinfo {author} {\bibfnamefont {F.~J.}\ \bibnamefont
  {Cao}}\ and\ \bibinfo {author} {\bibfnamefont {M.}~\bibnamefont {Feito}},\
  }\href {\doibase 10.1103/PhysRevE.79.041118} {\bibfield  {journal} {\bibinfo
  {journal} {Phys. Rev. E}\ }\textbf {\bibinfo {volume} {79}},\ \bibinfo
  {pages} {041118} (\bibinfo {year} {2009})}\BibitemShut {NoStop}%
\bibitem [{\citenamefont {Gaspard}(2013)}]{Gaspard_2013}%
  \BibitemOpen
  \bibfield  {author} {\bibinfo {author} {\bibfnamefont {P.}~\bibnamefont
  {Gaspard}},\ }\href {\doibase 10.1088/1367-2630/15/11/115014} {\bibfield
  {journal} {\bibinfo  {journal} {New Journal of Physics}\ }\textbf {\bibinfo
  {volume} {15}},\ \bibinfo {pages} {115014} (\bibinfo {year}
  {2013})}\BibitemShut {NoStop}%
\bibitem [{\citenamefont {Barato}\ and\ \citenamefont
  {Seifert}(2014{\natexlab{a}})}]{PhysRevLett_Barato}%
  \BibitemOpen
  \bibfield  {author} {\bibinfo {author} {\bibfnamefont {A.~C.}\ \bibnamefont
  {Barato}}\ and\ \bibinfo {author} {\bibfnamefont {U.}~\bibnamefont
  {Seifert}},\ }\href {\doibase 10.1103/PhysRevLett.112.090601} {\bibfield
  {journal} {\bibinfo  {journal} {Phys. Rev. Lett.}\ }\textbf {\bibinfo
  {volume} {112}},\ \bibinfo {pages} {090601} (\bibinfo {year}
  {2014}{\natexlab{a}})}\BibitemShut {NoStop}%
\bibitem [{\citenamefont {Barato}\ and\ \citenamefont
  {Seifert}(2014{\natexlab{b}})}]{PhysRevE_Barato}%
  \BibitemOpen
  \bibfield  {author} {\bibinfo {author} {\bibfnamefont {A.~C.}\ \bibnamefont
  {Barato}}\ and\ \bibinfo {author} {\bibfnamefont {U.}~\bibnamefont
  {Seifert}},\ }\href {\doibase 10.1103/PhysRevE.90.042150} {\bibfield
  {journal} {\bibinfo  {journal} {Phys. Rev. E}\ }\textbf {\bibinfo {volume}
  {90}},\ \bibinfo {pages} {042150} (\bibinfo {year}
  {2014}{\natexlab{b}})}\BibitemShut {NoStop}%
\bibitem [{\citenamefont {Horowitz}\ and\ \citenamefont
  {Esposito}(2014)}]{esposito2014prx}%
  \BibitemOpen
  \bibfield  {author} {\bibinfo {author} {\bibfnamefont {J.~M.}\ \bibnamefont
  {Horowitz}}\ and\ \bibinfo {author} {\bibfnamefont {M.}~\bibnamefont
  {Esposito}},\ }\href@noop {} {\bibfield  {journal} {\bibinfo  {journal}
  {Physical Review X}\ }\textbf {\bibinfo {volume} {4}},\ \bibinfo {pages}
  {031015} (\bibinfo {year} {2014})}\BibitemShut {NoStop}%
\bibitem [{\citenamefont {Rosinberg}\ and\ \citenamefont
  {Horowitz}(2016)}]{Rosinberg_2016}%
  \BibitemOpen
  \bibfield  {author} {\bibinfo {author} {\bibfnamefont {M.~L.}\ \bibnamefont
  {Rosinberg}}\ and\ \bibinfo {author} {\bibfnamefont {J.~M.}\ \bibnamefont
  {Horowitz}},\ }\href {\doibase 10.1209/0295-5075/116/10007} {\bibfield
  {journal} {\bibinfo  {journal} {{EPL} (Europhysics Letters)}\ }\textbf
  {\bibinfo {volume} {116}},\ \bibinfo {pages} {10007} (\bibinfo {year}
  {2016})}\BibitemShut {NoStop}%
\bibitem [{\citenamefont {Sagawa}\ and\ \citenamefont {Ueda}(2010)}]{sagawa}%
  \BibitemOpen
  \bibfield  {author} {\bibinfo {author} {\bibfnamefont {T.}~\bibnamefont
  {Sagawa}}\ and\ \bibinfo {author} {\bibfnamefont {M.}~\bibnamefont {Ueda}},\
  }\href@noop {} {\bibfield  {journal} {\bibinfo  {journal} {Phys. Rev. Lett.}\
  }\textbf {\bibinfo {volume} {104}},\ \bibinfo {pages} {090602} (\bibinfo
  {year} {2010})}\BibitemShut {NoStop}%
\bibitem [{\citenamefont {Sagawa}\ and\ \citenamefont {Ueda}(2012)}]{sagawa2}%
  \BibitemOpen
  \bibfield  {author} {\bibinfo {author} {\bibfnamefont {T.}~\bibnamefont
  {Sagawa}}\ and\ \bibinfo {author} {\bibfnamefont {M.}~\bibnamefont {Ueda}},\
  }\href@noop {} {\bibfield  {journal} {\bibinfo  {journal} {Phys. Rev. Lett.}\
  }\textbf {\bibinfo {volume} {109}},\ \bibinfo {pages} {180602} (\bibinfo
  {year} {2012})}\BibitemShut {NoStop}%
\bibitem [{\citenamefont {Jinwoo}(2019)}]{jinwoo2019fluctuation}%
  \BibitemOpen
  \bibfield  {author} {\bibinfo {author} {\bibfnamefont {L.}~\bibnamefont
  {Jinwoo}},\ }\href@noop {} {\bibfield  {journal} {\bibinfo  {journal}
  {Symmetry}\ }\textbf {\bibinfo {volume} {11}},\ \bibinfo {pages} {433}
  (\bibinfo {year} {2019})}\BibitemShut {NoStop}%
\bibitem [{\citenamefont {Jinwoo}\ and\ \citenamefont {Tanaka}(2015)}]{local}%
  \BibitemOpen
  \bibfield  {author} {\bibinfo {author} {\bibfnamefont {L.}~\bibnamefont
  {Jinwoo}}\ and\ \bibinfo {author} {\bibfnamefont {H.}~\bibnamefont
  {Tanaka}},\ }\href@noop {} {\bibfield  {journal} {\bibinfo  {journal}
  {Sci.Rep.}\ }\textbf {\bibinfo {volume} {5}},\ \bibinfo {pages} {7832}
  (\bibinfo {year} {2015})}\BibitemShut {NoStop}%
\bibitem [{\citenamefont {Jinwoo}\ and\ \citenamefont
  {Tanaka}(2014)}]{arxiv_v1}%
  \BibitemOpen
  \bibfield  {author} {\bibinfo {author} {\bibfnamefont {L.}~\bibnamefont
  {Jinwoo}}\ and\ \bibinfo {author} {\bibfnamefont {H.}~\bibnamefont
  {Tanaka}},\ }\href@noop {} {\bibfield  {journal} {\bibinfo  {journal}
  {arXiv:1403.1662v1}\ } (\bibinfo {year} {2014})}\BibitemShut {NoStop}%
\bibitem [{\citenamefont {Jarzynski}(2011)}]{jarReview}%
  \BibitemOpen
  \bibfield  {author} {\bibinfo {author} {\bibfnamefont {C.}~\bibnamefont
  {Jarzynski}},\ }\href@noop {} {\bibfield  {journal} {\bibinfo  {journal}
  {Annu. Rev. Codens. Matter Phys.}\ }\textbf {\bibinfo {volume} {2}},\
  \bibinfo {pages} {329} (\bibinfo {year} {2011})}\BibitemShut {NoStop}%
\bibitem [{\citenamefont {Seifert}(2012)}]{revSeifert}%
  \BibitemOpen
  \bibfield  {author} {\bibinfo {author} {\bibfnamefont {U.}~\bibnamefont
  {Seifert}},\ }\href@noop {} {\bibfield  {journal} {\bibinfo  {journal} {Rep.
  Prog. Phys.}\ }\textbf {\bibinfo {volume} {75}},\ \bibinfo {pages} {126001}
  (\bibinfo {year} {2012})}\BibitemShut {NoStop}%
\bibitem [{\citenamefont {Spinney}\ and\ \citenamefont {Ford}(2013)}]{review}%
  \BibitemOpen
  \bibfield  {author} {\bibinfo {author} {\bibfnamefont {R.}~\bibnamefont
  {Spinney}}\ and\ \bibinfo {author} {\bibfnamefont {I.}~\bibnamefont {Ford}},\
  }\enquote {\bibinfo {title} {Fluctuation relations: A pedagogical
  overview},}\ in\ \href@noop {} {\emph {\bibinfo {booktitle} {Nonequilibrium
  Statistical Physics of Small Systems}}}\ (\bibinfo  {publisher} {Wiley-VCH
  Verlag GmbH \& Co. KGaA},\ \bibinfo {year} {2013})\ pp.\ \bibinfo {pages}
  {3--56}\BibitemShut {NoStop}%
\bibitem [{\citenamefont {Cover}\ and\ \citenamefont {Thomas}(2012)}]{cover}%
  \BibitemOpen
  \bibfield  {author} {\bibinfo {author} {\bibfnamefont {T.~M.}\ \bibnamefont
  {Cover}}\ and\ \bibinfo {author} {\bibfnamefont {J.~A.}\ \bibnamefont
  {Thomas}},\ }\href@noop {} {\emph {\bibinfo {title} {Elements of information
  theory}}}\ (\bibinfo  {publisher} {John Wiley \& Sons},\ \bibinfo {year}
  {2012})\BibitemShut {NoStop}%
\bibitem [{\citenamefont {Ponmurugan}(2010)}]{ponmurugan2010}%
  \BibitemOpen
  \bibfield  {author} {\bibinfo {author} {\bibfnamefont {M.}~\bibnamefont
  {Ponmurugan}},\ }\href@noop {} {\bibfield  {journal} {\bibinfo  {journal}
  {Physical Review E}\ }\textbf {\bibinfo {volume} {82}},\ \bibinfo {pages}
  {031129} (\bibinfo {year} {2010})}\BibitemShut {NoStop}%
\bibitem [{\citenamefont {Horowitz}\ and\ \citenamefont
  {Vaikuntanathan}(2010)}]{horowitz2010}%
  \BibitemOpen
  \bibfield  {author} {\bibinfo {author} {\bibfnamefont {J.~M.}\ \bibnamefont
  {Horowitz}}\ and\ \bibinfo {author} {\bibfnamefont {S.}~\bibnamefont
  {Vaikuntanathan}},\ }\href@noop {} {\bibfield  {journal} {\bibinfo  {journal}
  {Physical Review E}\ }\textbf {\bibinfo {volume} {82}},\ \bibinfo {pages}
  {061120} (\bibinfo {year} {2010})}\BibitemShut {NoStop}%
\bibitem [{\citenamefont {Kurchan}(1998)}]{kur}%
  \BibitemOpen
  \bibfield  {author} {\bibinfo {author} {\bibfnamefont {J.}~\bibnamefont
  {Kurchan}},\ }\href@noop {} {\bibfield  {journal} {\bibinfo  {journal}
  {Journal of Physics A: Mathematical and General}\ }\textbf {\bibinfo {volume}
  {31}},\ \bibinfo {pages} {3719} (\bibinfo {year} {1998})}\BibitemShut
  {NoStop}%
\bibitem [{\citenamefont {Maes}(1999)}]{maes1999}%
  \BibitemOpen
  \bibfield  {author} {\bibinfo {author} {\bibfnamefont {C.}~\bibnamefont
  {Maes}},\ }\href@noop {} {\bibfield  {journal} {\bibinfo  {journal} {Journal
  of statistical physics}\ }\textbf {\bibinfo {volume} {95}},\ \bibinfo {pages}
  {367} (\bibinfo {year} {1999})}\BibitemShut {NoStop}%
\bibitem [{\citenamefont {Jarzynski}(2000)}]{jar2000}%
  \BibitemOpen
  \bibfield  {author} {\bibinfo {author} {\bibfnamefont {C.}~\bibnamefont
  {Jarzynski}},\ }\href@noop {} {\bibfield  {journal} {\bibinfo  {journal} {J.
  Stat. Phys.}\ }\textbf {\bibinfo {volume} {98}},\ \bibinfo {pages} {77}
  (\bibinfo {year} {2000})}\BibitemShut {NoStop}%
\bibitem [{\citenamefont {Goldstein}\ \emph {et~al.}(2001)\citenamefont
  {Goldstein}, \citenamefont {Jr.},\ and\ \citenamefont {Safko}}]{goldstein}%
  \BibitemOpen
  \bibfield  {author} {\bibinfo {author} {\bibfnamefont {H.}~\bibnamefont
  {Goldstein}}, \bibinfo {author} {\bibfnamefont {C.~P.~P.}\ \bibnamefont
  {Jr.}}, \ and\ \bibinfo {author} {\bibfnamefont {J.~L.}\ \bibnamefont
  {Safko}},\ }\href@noop {} {\emph {\bibinfo {title} {Classical Mechanics (3rd
  Edition)}}}\ (\bibinfo  {publisher} {Pearson},\ \bibinfo {year}
  {2001})\BibitemShut {NoStop}%
\bibitem [{\citenamefont {Parrondo}\ \emph {et~al.}(2015)\citenamefont
  {Parrondo}, \citenamefont {Horowitz},\ and\ \citenamefont
  {Sagawa}}]{parrondo2015}%
  \BibitemOpen
  \bibfield  {author} {\bibinfo {author} {\bibfnamefont {J.~M.}\ \bibnamefont
  {Parrondo}}, \bibinfo {author} {\bibfnamefont {J.~M.}\ \bibnamefont
  {Horowitz}}, \ and\ \bibinfo {author} {\bibfnamefont {T.}~\bibnamefont
  {Sagawa}},\ }\href@noop {} {\bibfield  {journal} {\bibinfo  {journal} {Nature
  physics}\ }\textbf {\bibinfo {volume} {11}},\ \bibinfo {pages} {131}
  (\bibinfo {year} {2015})}\BibitemShut {NoStop}%
\bibitem [{\citenamefont {Kawai}\ \emph {et~al.}(2007)\citenamefont {Kawai},
  \citenamefont {Parrondo},\ and\ \citenamefont {den Broeck}}]{kawai}%
  \BibitemOpen
  \bibfield  {author} {\bibinfo {author} {\bibfnamefont {R.}~\bibnamefont
  {Kawai}}, \bibinfo {author} {\bibfnamefont {J.~M.~R.}\ \bibnamefont
  {Parrondo}}, \ and\ \bibinfo {author} {\bibfnamefont {C.~V.}\ \bibnamefont
  {den Broeck}},\ }\href@noop {} {\bibfield  {journal} {\bibinfo  {journal}
  {Phys. Rev. Lett.}\ }\textbf {\bibinfo {volume} {98}},\ \bibinfo {pages}
  {080602} (\bibinfo {year} {2007})}\BibitemShut {NoStop}%
\bibitem [{tak(2010)}]{takara}%
  \BibitemOpen
  \href@noop {} {\bibfield  {journal} {\bibinfo  {journal} {Physics Letters A}\
  }\textbf {\bibinfo {volume} {375}},\ \bibinfo {pages} {88 } (\bibinfo {year}
  {2010})}\BibitemShut {NoStop}%
\bibitem [{has(2010)}]{hasegawa}%
  \BibitemOpen
  \href@noop {} {\bibfield  {journal} {\bibinfo  {journal} {Physics Letters A}\
  }\textbf {\bibinfo {volume} {374}},\ \bibinfo {pages} {1001 } (\bibinfo
  {year} {2010})}\BibitemShut {NoStop}%
\bibitem [{\citenamefont {Esposito}\ and\ \citenamefont {Van~den
  Broeck}(2011)}]{esposito2011}%
  \BibitemOpen
  \bibfield  {author} {\bibinfo {author} {\bibfnamefont {M.}~\bibnamefont
  {Esposito}}\ and\ \bibinfo {author} {\bibfnamefont {C.}~\bibnamefont {Van~den
  Broeck}},\ }\href@noop {} {\bibfield  {journal} {\bibinfo  {journal}
  {Europhys. Lett.}\ }\textbf {\bibinfo {volume} {95}},\ \bibinfo {pages}
  {40004} (\bibinfo {year} {2011})}\BibitemShut {NoStop}%
\bibitem [{\citenamefont {McGrath}\ \emph {et~al.}(2017)\citenamefont
  {McGrath}, \citenamefont {Jones}, \citenamefont {ten Wolde},\ and\
  \citenamefont {Ouldridge}}]{Thomas2017}%
  \BibitemOpen
  \bibfield  {author} {\bibinfo {author} {\bibfnamefont {T.}~\bibnamefont
  {McGrath}}, \bibinfo {author} {\bibfnamefont {N.~S.}\ \bibnamefont {Jones}},
  \bibinfo {author} {\bibfnamefont {P.~R.}\ \bibnamefont {ten Wolde}}, \ and\
  \bibinfo {author} {\bibfnamefont {T.~E.}\ \bibnamefont {Ouldridge}},\ }\href
  {\doibase 10.1103/PhysRevLett.118.028101} {\bibfield  {journal} {\bibinfo
  {journal} {Phys. Rev. Lett.}\ }\textbf {\bibinfo {volume} {118}},\ \bibinfo
  {pages} {028101} (\bibinfo {year} {2017})}\BibitemShut {NoStop}%
\bibitem [{\citenamefont {Becker}\ \emph {et~al.}(2015)\citenamefont {Becker},
  \citenamefont {Mugler},\ and\ \citenamefont {ten Wolde}}]{Becker2015}%
  \BibitemOpen
  \bibfield  {author} {\bibinfo {author} {\bibfnamefont {N.~B.}\ \bibnamefont
  {Becker}}, \bibinfo {author} {\bibfnamefont {A.}~\bibnamefont {Mugler}}, \
  and\ \bibinfo {author} {\bibfnamefont {P.~R.}\ \bibnamefont {ten Wolde}},\
  }\href {\doibase 10.1103/PhysRevLett.115.258103} {\bibfield  {journal}
  {\bibinfo  {journal} {Phys. Rev. Lett.}\ }\textbf {\bibinfo {volume} {115}},\
  \bibinfo {pages} {258103} (\bibinfo {year} {2015})}\BibitemShut {NoStop}%
\bibitem [{\citenamefont {Ouldridge}\ \emph {et~al.}(2017)\citenamefont
  {Ouldridge}, \citenamefont {Govern},\ and\ \citenamefont {ten
  Wolde}}]{Ouldridge2017}%
  \BibitemOpen
  \bibfield  {author} {\bibinfo {author} {\bibfnamefont {T.~E.}\ \bibnamefont
  {Ouldridge}}, \bibinfo {author} {\bibfnamefont {C.~C.}\ \bibnamefont
  {Govern}}, \ and\ \bibinfo {author} {\bibfnamefont {P.~R.}\ \bibnamefont {ten
  Wolde}},\ }\href {\doibase 10.1103/PhysRevX.7.021004} {\bibfield  {journal}
  {\bibinfo  {journal} {Phys. Rev. X}\ }\textbf {\bibinfo {volume} {7}},\
  \bibinfo {pages} {021004} (\bibinfo {year} {2017})}\BibitemShut {NoStop}%
\bibitem [{\citenamefont {Tsai}\ and\ \citenamefont
  {Nussinov}(2014)}]{tsai2014unified}%
  \BibitemOpen
  \bibfield  {author} {\bibinfo {author} {\bibfnamefont {C.-J.}\ \bibnamefont
  {Tsai}}\ and\ \bibinfo {author} {\bibfnamefont {R.}~\bibnamefont
  {Nussinov}},\ }\href@noop {} {\bibfield  {journal} {\bibinfo  {journal} {PLoS
  computational biology}\ }\textbf {\bibinfo {volume} {10}},\ \bibinfo {pages}
  {e1003394} (\bibinfo {year} {2014})}\BibitemShut {NoStop}%
\bibitem [{\citenamefont {Cuendet}\ \emph {et~al.}(2016)\citenamefont
  {Cuendet}, \citenamefont {Weinstein},\ and\ \citenamefont
  {LeVine}}]{cuendet2016allostery}%
  \BibitemOpen
  \bibfield  {author} {\bibinfo {author} {\bibfnamefont {M.~A.}\ \bibnamefont
  {Cuendet}}, \bibinfo {author} {\bibfnamefont {H.}~\bibnamefont {Weinstein}},
  \ and\ \bibinfo {author} {\bibfnamefont {M.~V.}\ \bibnamefont {LeVine}},\
  }\href@noop {} {\bibfield  {journal} {\bibinfo  {journal} {Journal of
  chemical theory and computation}\ }\textbf {\bibinfo {volume} {12}},\
  \bibinfo {pages} {5758} (\bibinfo {year} {2016})}\BibitemShut {NoStop}%
\end{thebibliography}%

\end{document}